%% file: colm2026_conference.tex
\documentclass{article}
\usepackage[final]{colm2026_conference}
\usepackage{microtype}
\usepackage{hyperref}
\usepackage{url}
\usepackage{booktabs}
\usepackage{graphicx}
\usepackage{amsmath}
\usepackage{amssymb}
\usepackage{multirow}
\usepackage{xcolor}
\usepackage{subcaption}
\usepackage{lineno}

\title{Heard but Not Heeded: Paralinguistic Information \\ Encoding and Loss in Audio-Language Models}
\author{
Bhuvan Koduru \\
Language Technologies Institute \\
Carnegie Mellon University \\
\texttt{bkoduru@andrew.cmu.edu}
\And
Dareen Safar B Alharthi \\
Language Technologies Institute \\
Carnegie Mellon University \\
\texttt{dalharth@andrew.cmu.edu}
\And
Rita Singh \\
Language Technologies Institute \\
Carnegie Mellon University \\
\texttt{rsingh@cs.cmu.edu}
\And
Bhiksha Raj \\
Language Technologies Institute \\
Carnegie Mellon University \\
\texttt{bhiksha@cs.cmu.edu} }

\begin{document}

\ifcolmsubmission
\linenumbers
\fi

\maketitle

\input{sections/abstract}

\input{sections/introduction}

\input{sections/related_work}
\input{sections/method}
\input{sections/exp}
\input{sections/results}

\input{sections/discussion}

\input{sections/conclusion}

\section*{LLM Usage Disclosure}
Claude (Anthropic) was used for research discussion and writing assistance. All experiments, code, results, and final scientific conclusions are the authors' own.

\newpage
\bibliography{references}
\bibliographystyle{colm2026_conference}

\newpage
\input{sections/appendix}
\end{document}

%% file: sections/abstract.tex
\begin{abstract}
Audio language models are designed to understand speech, yet it remains unclear whether they capture how something is said beyond what is said. We present a mechanistic analysis of paralinguistic information in four open source models, Whisper-large-v2, Qwen2-Audio-7B Instruct, Qwen2.5-Omni-7B, and Chroma-4B, using the Expresso dataset with controlled speaking styles. We combine centered kernel alignment, linear probing with leave one speaker out evaluation, open ended tone prediction, and a content prosody leakage metric to trace how style information moves from the audio encoder to the final output. All models strongly encode speaking style in the late encoder, that is, the top third of the audio encoder's layers, but this information is consistently degraded before reaching the output. The projector reshapes representation geometry without removing information, while decoders differ in how much style they preserve depending on architecture and training objective. At the output level, models fall into two behaviors. Some are content driven, where predictions depend mainly on text. Others are acoustic driven, where predictions vary with speaking style. The leakage metric quantifies this difference, and qualitative results confirm it. Overall, we identify a gap between what models encode and what they use, highlighting a key limitation in current audio language models.
\end{abstract}

%% file: sections/introduction.tex
\section{Introduction}

Recent advances in large language models (LLMs) have expanded their capabilities beyond text to include multimodal understanding and generation \citep{xu2025qwen25omni,ghosh2024gama,deshmukh2023pengi,deshmukh2025mellow,chen2026chroma,peng2023kosmos}. Modern systems can process speech, audio, and visual inputs, and some can respond across modalities \citep{ge2024seed,pan2023kosmos}, for example, generating spoken responses or producing images alongside text. These multimodal language models promise more natural and expressive interaction. However, despite this progress, their performance still lags behind text only models in many settings, particularly when it comes to understanding information that goes beyond literal content.

Speech carries much more than just content. When someone whispers, laughs, hesitates, or changes their tone, they communicate additional meaning through how they speak, not just what they say. This includes prosody, emotion, speaking rate, and voice quality, which we refer to as paralinguistic information. Beyond conveying intent and tone, speech also encodes rich biometric signals. A person’s voice can reveal characteristics such as identity, age, gender, accent, and emotional state, and can even provide signals about health and mental status \citep{singh2019profiling}. This layer of information is essential in human communication and is widely used in real world systems, including speaker recognition \citep{teixeira2022towards}, emotion detection \citep{dhamyal2024prompting,dhamyal2022positional}, and personalization \citep{joseph2024speaker}. As audio language models are deployed in voice assistants and conversational systems, an important question arises: do these models actually capture and use this information, or do they ignore it in favor of the spoken content?

Most modern ALMs follow a simple pipeline. An audio encoder processes the raw waveform, a projector maps it into the language model space, and a decoder generates text \citep{chen2026chroma,xu2025qwen25omni,chu2024qwen2audio}. While this setup works well for tasks like speech recognition, it is not clear whether it preserves paralinguistic information or whether this information is lost because the model is mainly trained to recover text. Prior work has studied prosodic and acoustic representations in standalone speech encoders \citep{yang2021superb, sicherman2023analysing}. However, these studies do not track how this information moves through the full encoder, projector, and decoder pipeline of modern audio language models. They also do not explain why models sometimes appear insensitive to tone, emotion, or speaking style at the output level. This leaves an important gap in understanding how these models process, and sometimes ignore, paralinguistic signals.

In this paper, we present a comprehensive empirical analysis of how paralinguistic information is encoded and lost in audio language models. We study four representative models and analyze how this information changes across their internal layers using a set of complementary methods. Our goal is to identify where and why paralinguistic information is lost or ignored, and to better understand the architectural and training choices that lead to this behavior. We focus specifically on speaking style (prosody, tone of delivery) as our paralinguistic dimension of interest; other acoustic attributes such as speaker identity or accent are outside the scope of this study. Our contributions are as follows:
\begin{itemize}
    \item We design a four part analysis framework, including centered kernel alignment (CKA) \citep{kornblith2019similarity}, linear probing, open ended tone evaluation, and content prosody leakage, and apply it across all stages of four models.

    \item We find that all models encode paralinguistic information strongly in the audio encoder, but differ in how they use it later.

    \item We show that the projector changes the geometry of representations without removing the information itself, suggesting that it reshapes the signal rather than discarding it.

    \item We identify two types of behavior: content driven models, where tone predictions depend mostly on the words, and acoustic driven models, where predictions reflect how the speech actually sounds.

    \item We introduce a new metric that measures how much model predictions depend on content versus acoustic style.
\end{itemize}

Overall, our results suggest that these models hear paralinguistic information but often do not use it. We discuss why this happens and outline directions for building models that better capture the full meaning of speech.

%% file: sections/related_work.tex
\section{Related Work}

Speech representations have been widely studied in tasks that rely on information beyond content, such as emotion recognition, speaker traits, and prosody \citep{pasad2021layer, baevski2020wav2vec,chen2022wavlm,yang2021superb}. Benchmarks like SUPERB \citep{yang2021superb} show that modern speech encoders capture rich paralinguistic information and perform well on tasks such as emotion classification. More focused studies further confirm that signals like pitch, speaking style, and prosody are present in self-supervised speech representations \citep{sicherman2023analysing}. However, these works mainly study speech encoders in isolation and do not explain how this information behaves once it is passed into larger systems.

To understand what models encode, prior work has introduced tools such as centered kernel alignment \citep{kornblith2019similarity}, which compares representations across layers, and linear probing \citep{alain2016understanding}, which measures whether specific information can be decoded from representations. These methods have been applied to vision models, diffusion models \citep{surkov2024unpacking}, and language models \citep{bricken2023monosemanticity}, providing insights into how information is structured inside deep networks. However, similar analyses for audio and speech language models remain limited, especially when it comes to tracking how information moves from the audio input to the final output.

In this work, we bring these directions together. Instead of studying speech encoders alone or evaluating models only at the output level, we trace paralinguistic information through the full pipeline, from the audio encoder to the final prediction. This allows us to identify where the information is preserved, where it changes, and where it is ultimately ignored.

%% file: sections/method.tex
\section{Paralinguistic Analysis Framework}
We study how paralinguistic information is represented and used in audio language models by analyzing both internal representations and model outputs. Our analysis consists of three components: representation similarity, information accessibility, and output-level behavior.

For each audio input, we extract hidden states from all layers using a single forward pass and apply mean pooling over time to obtain fixed-size representations. Mean pooling is applied uniformly across all four models, which keeps the comparison controlled even though it discards temporal dynamics that may matter for styles such as laughing (see Section~\ref{sec:limitations}); the consistent probing peaks of $\sim$82--85\% we observe across all models (Section~\ref{sec:results}) indicate that a substantial part of the style signal survives pooling. For models with interleaved text and audio tokens, we extract representations only at audio token positions to avoid mixing modalities.

We first analyze representation geometry using centered kernel alignment (CKA) \citep{kornblith2019similarity}. For a pair of inputs $(a_1, a_2)$, let $X, Y \in \mathbb{R}^{n \times d}$ be their layer-wise representation matrices (rows indexed by the $n$ paired examples used for aggregation) and let $K = XX^\top$, $L = YY^\top$ be the corresponding Gram matrices. The Hilbert-Schmidt Independence Criterion (HSIC) between two $n\times n$ Gram matrices $K, L$ is
\begin{equation}
\mathrm{HSIC}(K,L) = \frac{1}{(n-1)^2}\,\mathrm{tr}(KHLH), \qquad H = I_n - \tfrac{1}{n}\mathbf{1}\mathbf{1}^\top,
\end{equation}
where $H$ is the centering matrix. Linear CKA is then the normalized HSIC between the two Gram matrices:
\begin{equation}
\mathrm{CKA}(X,Y) = \frac{\mathrm{HSIC}(K,L)}
{\sqrt{\mathrm{HSIC}(K,K)\,\mathrm{HSIC}(L,L)}} \in [0,1].
\end{equation}
We mean-center $X$ and $Y$ column-wise before computing $K$ and $L$, so $HKH = K$ and $HLH = L$ and the expression reduces to $\mathrm{CKA}(X,Y) = \|X^\top Y\|_F^2 / (\|X^\top X\|_F \, \|Y^\top Y\|_F)$. High CKA indicates that the model maps the inputs to similar representations, while low CKA indicates stronger separation.

To measure whether paralinguistic information is encoded in the representations, we use linear probing. A classifier is trained on representations from each layer to predict speaking style. This allows us to test whether style information is present even when representations appear geometrically similar. We then evaluate whether models use this information at the output level. Each model is prompted to describe the tone of speech in natural language, and responses are mapped to predefined tone categories.

Finally, we quantify whether predictions are driven by semantic content or acoustic style. Let $x$ denote the text content of an utterance and $e$ its ground-truth speaking-style label. For an audio clip $a$ built from $(x, e)$, let $\hat{t} = f(a)$ denote the tone bucket predicted by model $f$ from the clip's audio. We compute the normalized mutual information (NMI) between predicted tone and text content, $\mathrm{NMI}(\hat{t}, x)$, and between predicted tone and speaking style, $\mathrm{NMI}(\hat{t}, e)$, each measured over the full set of clips. We define the leakage ratio as
\begin{equation}
\frac{\mathrm{NMI}(\hat{t}, x)}{\mathrm{NMI}(\hat{t}, x) + \mathrm{NMI}(\hat{t}, e)},
\end{equation}

which measures how much the prediction depends on content versus speaking style. Values close to 1 mean the prediction relies mostly on content, while values close to 0 mean it relies more on acoustic style. Because $x$ ranges over many more distinct values (one per unique utterance) than $e$ (7 style labels), content has substantially higher entropy than style; NMI normalization partially controls for this, but the resulting leakage ratios should be read as relative across models rather than as absolute probabilities (see Section~\ref{sec:limitations}).

%% file: sections/exp.tex
\section{Experimental Setup}

\textbf{Model selection.} We analyze four models spanning the major axes of ALM design, so differences in paralinguistic preservation can be attributed to specific architectural or training choices. Whisper-large-v2 \citep{radford2023robust} is encoder-only (no LM decoder), a baseline for a purely text-recovery objective. Qwen2-Audio-7B-Instruct \citep{chu2024qwen2audio} and Qwen2.5-Omni-7B \citep{xu2025qwen25omni} share a nearly identical Whisper-derived encoder and Qwen LLM backbone but differ in objective: text-only vs.\ additionally speech-generating, isolating the training-objective effect while holding architecture fixed. Chroma-4B \citep{chen2026chroma} adds a modular reasoning ("thinker") stage plus voice-cloning synthesis, testing whether a synthesis objective preserves style differently from Omni's joint objective. Table~\ref{tab:models} summarizes their architectures.

\begin{table}[t]
\centering
\caption{Summary of the four models analyzed. Layer counts are for the components probed in this study (see Appendix~\ref{app:layers} for full layer accounting).}
\label{tab:models}
\resizebox{\textwidth}{!}{
\begin{tabular}{lllll}
\toprule
\textbf{Model} & \textbf{Encoder} & \textbf{Projector} & \textbf{Decoder} & \textbf{Training objective} \\
\midrule
Whisper-large-v2 & 32-layer Whisper encoder & — & — & ASR (text transcription only) \\
Qwen2-Audio-7B-Instruct & 33-layer Whisper-derived & Linear (1280$\to$4096) & 33-layer Qwen-7B LLM & Speech-to-text instruction tuning \\
Qwen2.5-Omni-7B & 32-layer Whisper-derived & Multi-layer MLP & 28-layer Qwen LLM + TTS & Joint text and speech generation \\
Chroma-4B & 32-layer thinker audio encoder & — (shared backbone) & Llama3 backbone + 4-layer decoder & Reasoning + voice-cloning synthesis \\
\bottomrule
\end{tabular}
}
\end{table}

Throughout the paper, we use \emph{late encoder} to refer to the top third of an audio encoder's layers (e.g., layers 20--31 of Whisper-large-v2's 32-layer encoder), as opposed to early or middle encoder layers.

\textbf{Dataset.} We use the Expresso dataset \citep{nguyen2023expresso}, which contains the same utterances spoken in different styles. This allows us to isolate how something is said while keeping the content fixed. Let $x$ denote an utterance and $e \in \mathcal{E}$ a speaking style, where $\mathcal{E}$ is the set of 7 speaking-style labels used in our analysis (confused, default, enunciated, happy, laughing, sad, whisper), so $|\mathcal{E}| = 7$. Each clip is written as $a(x,e,s)$ for speaker $s$. We construct cross-style pairs by fixing the text and speaker while varying style:
\begin{equation}
\mathcal{P} = \left\{ \bigl(a(x,e_i,s), a(x,e_j,s)\bigr) \;\middle|\; e_i \neq e_j \in \mathcal{E} \right\}.
\end{equation}
Each pair differs only in paralinguistic delivery, enabling controlled evaluation of how models capture speaking style. Linear probes are trained with leave-one-speaker-out (LOSO) cross-validation (L2-regularized logistic regression, $C=0.1$, SAGA solver, features standardized per fold); full hyperparameters are given in Appendix~\ref{app:probe}.

%% file: sections/results.tex
\section{Results} \label{sec:results}

\subsection{Whisper: A Style-Encoding ASR Baseline}

Figure~\ref{fig:cka_all} shows Whisper-large-v2 CKA: all style pairs remain tightly clustered between 0.92 and 1.00 throughout all 32 encoder layers, converging to $\sim$0.99 at the final layer. This geometric near-identity is consistent with Whisper's ASR objective. However, Figure~\ref{fig:probe_cross} reveals a striking dissociation: linear probing accuracy rises monotonically from $\sim$51\% at layer 0 to $\sim$83\% at layer 31. Whisper \emph{does} encode speaking style in a linearly accessible form — its representations are geometrically similar across styles, but style information is present. This CKA-probing dissociation establishes a key methodological point: high CKA does not imply absence of information, only geometric similarity. Whisper is a \emph{style-encoding, geometry-collapsing} baseline.

\subsection{Encoder Encoding: All Models Peak at Late Encoder}

\begin{figure}[t]
    \centering
    \includegraphics[width=0.95\textwidth]{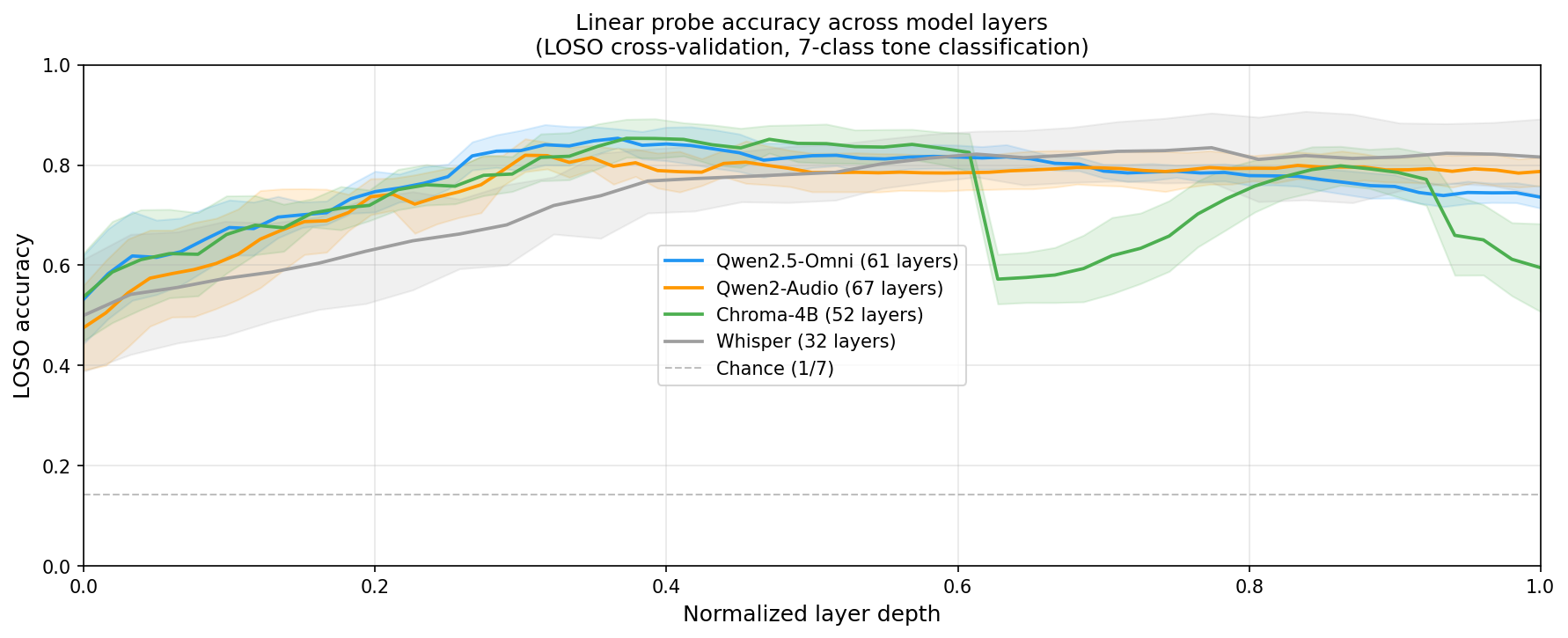}
    \caption{Linear probe accuracy (LOSO, 7-class) across all four models on normalized layer depth (0 = first encoder layer, 1 = final decoder layer; see Appendix~\ref{app:layers} for the encoder/projector/decoder boundaries per model). All models peak in the late audio encoder ($\sim$82--85\%). Qwen2-Audio (orange) maintains flat decoder accuracy. Qwen2.5-Omni (blue) declines gradually. Chroma (green) drops sharply at the backbone boundary then partially recovers. Whisper (gray) rises monotonically. Chance = 14.3\% (dashed).}
    \label{fig:probe_cross}
\end{figure}

Figure~\ref{fig:probe_cross} shows probing accuracy across all models on normalized layer depth. All four models show a consistent encoder trajectory: accuracy rises from $\sim$47--55\% at layer 0 to a peak of $\sim$82--85\% in the late encoder. Specifically, Qwen2-Audio peaks at $\sim$82\% around encoder layer 20, Qwen2.5-Omni peaks at $\sim$84\% around layer 22, and Chroma peaks at $\sim$85\% around layers 18--20. All peaks are well above the 14.3\% chance level.

The per-style-pair CKA trajectories (Figure~\ref{fig:cka_all} in Appendix~\ref{app:cka_fig}) corroborate this: inter-style CKA decreases substantially in the late encoder for all models. Chroma's thinker audio encoder shows the most dramatic behavior: CKA drops from $\sim$0.65 to near zero ($\sim$0.03--0.06) at layers 20--31, the largest geometric style separation of any model in our study. The convergent evidence from probing and CKA provides robust support: all models encode substantial paralinguistic information in the late audio encoder. To contextualize the $\sim$82--85\% probe accuracy against an established paralinguistic baseline, we note that wav2vec 2.0 fine-tuned for speech emotion recognition achieves roughly 70--80\% accuracy on comparably sized (7-class) emotion taxonomies \citep{baevski2020wav2vec, yang2021superb}; our encoder-probe accuracies are near or above this range, suggesting that late-encoder representations carry close to the linearly accessible ceiling for style information, though a controlled in-domain SER baseline on Expresso itself would sharpen this comparison.

\subsection{The Projector: Geometric Disruption Without Information Loss}

All three encoder-projector-decoder models show a sharp CKA discontinuity at the projector boundary. In Qwen2-Audio, CKA drops sharply from $\sim$0.95 to $\sim$0.60 at the projector. In Omni, the drop is more dramatic. Yet probing accuracy at the projector remains close to the encoder peak for both models ($\sim$78--81\%), indicating that the projector \emph{reorganizes} rather than \emph{destroys} paralinguistic information.

This CKA-probing dissociation is mechanistically informative. Qwen2-Audio's single linear projector maps from 1280-dim to 4096-dim space — a 3$\times$ dimensional expansion that dramatically alters representational geometry while preserving the underlying linear structure. This explains why CKA collapses but probing remains stable: a linear transformation changes geometry without destroying linear decodability.

\subsection{Decoder Divergence: Three Distinct Patterns}

\begin{figure}[t]
    \centering
    \begin{subfigure}[b]{0.48\textwidth}
        \includegraphics[width=\textwidth]{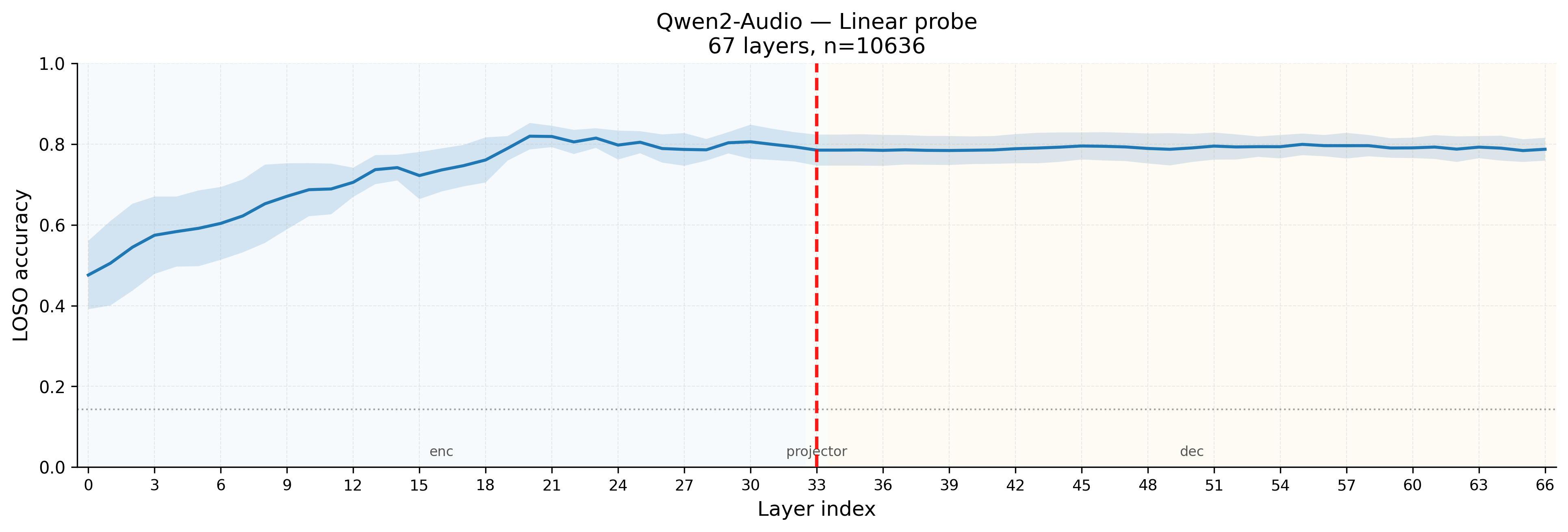}
        \caption{Qwen2-Audio-7B (67 layers)}
    \end{subfigure}
    \hfill
    \begin{subfigure}[b]{0.48\textwidth}
        \includegraphics[width=\textwidth]{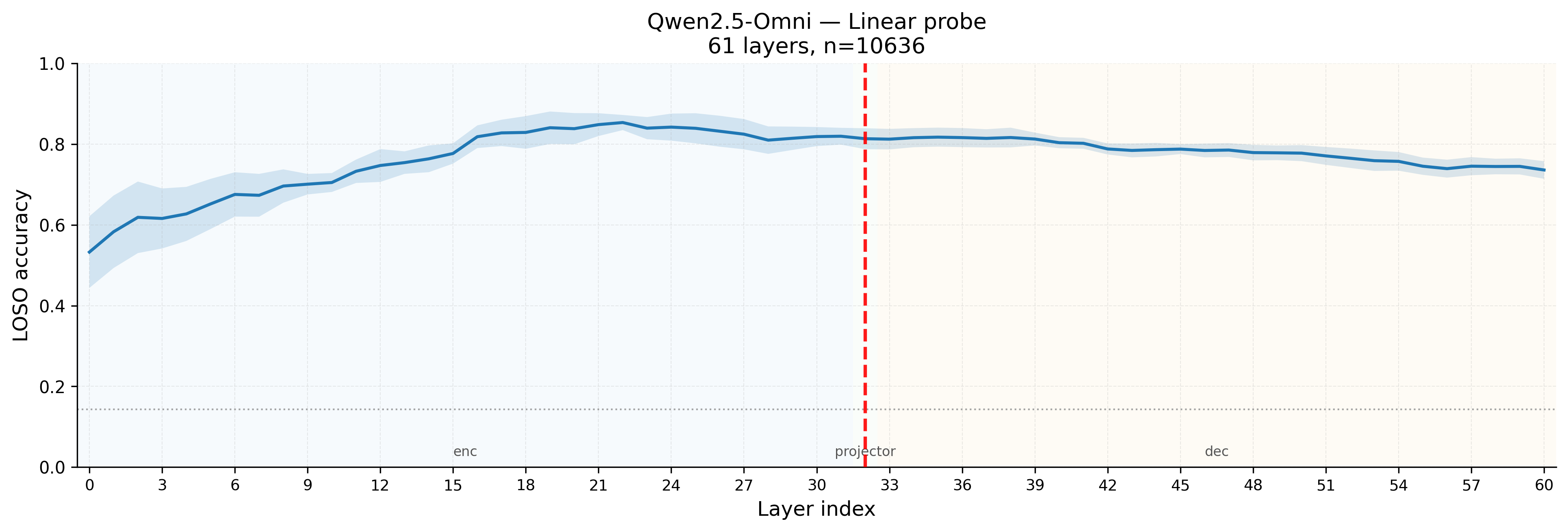}
        \caption{Qwen2.5-Omni-7B (61 layers)}
    \end{subfigure}
    \vspace{0.5em}
    \begin{subfigure}[b]{0.48\textwidth}
        \includegraphics[width=\textwidth]{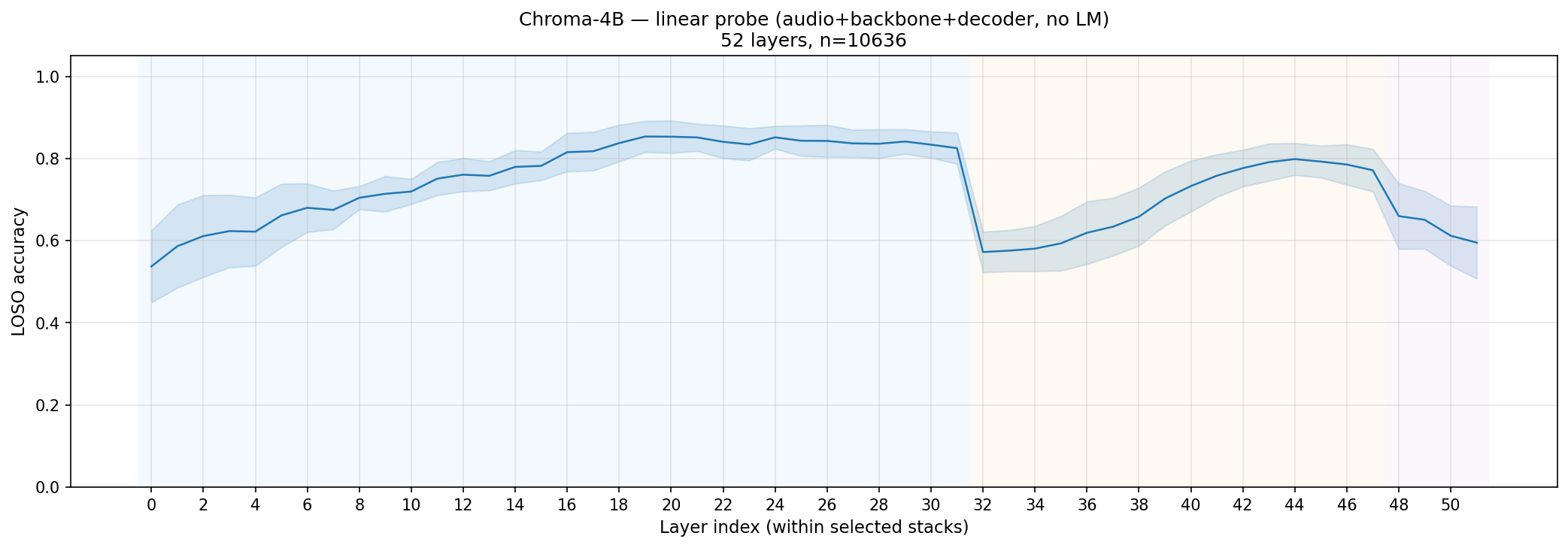}
        \caption{Chroma-4B (52 layers, no thinker LM)}
    \end{subfigure}
    \hfill
    \begin{subfigure}[b]{0.48\textwidth}
        \includegraphics[width=\textwidth]{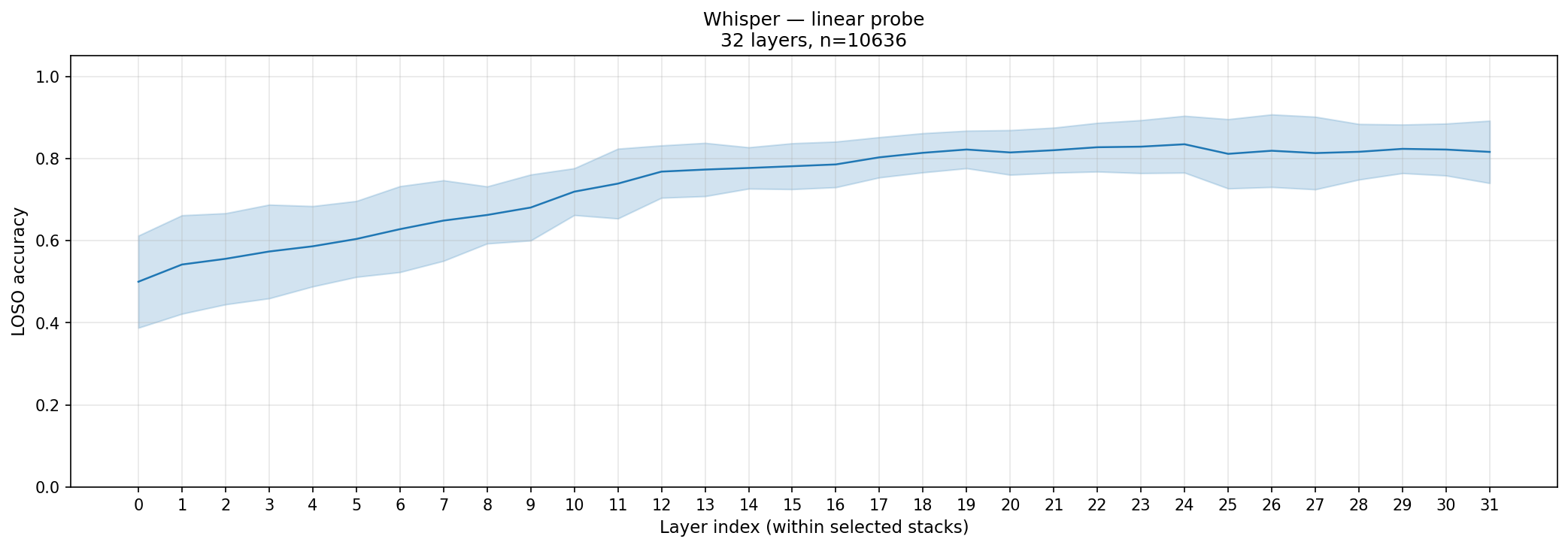}
        \caption{Whisper-large-v2 (32 layers)}
    \end{subfigure}
    \caption{Per-model linear probe accuracy (LOSO) on each model's own (unnormalized) layer index, with $\pm$1 std shading across folds. Whereas Figure~\ref{fig:probe_cross} overlays all four models on a shared normalized depth axis to compare cross-model trends, this figure shows each model's absolute layer-by-layer trajectory and fold variance individually.}
    \label{fig:probe_individual}
\end{figure}

The three models diverge sharply in their decoder behavior, as shown in Figures~\ref{fig:probe_cross} and \ref{fig:probe_individual}.

\textbf{Qwen2-Audio}: Probing accuracy remains flat at $\sim$78--80\% across all 33 decoder layers (Figure~\ref{fig:probe_individual}). Style information persists in a linearly accessible form throughout the LLM backbone. Simultaneously, inter-style CKA drops dramatically in the decoder — representations become geometrically highly distinct across styles. This probing-CKA dissociation indicates a non-isometric transformation: style information is reorganized geometrically while its linear decodability is preserved.

\textbf{Qwen2.5-Omni}: Probing accuracy declines gradually from $\sim$81\% at the projector to $\sim$74\% by the final decoder layer — a consistent, monotonic erosion of paralinguistic information. CKA declines in parallel, suggesting genuine information loss.

\textbf{Chroma}: The backbone reveals a distinctive two-phase pattern. Probing drops sharply from $\sim$85\% at the audio encoder peak to $\sim$58\% at backbone entry (layer 32), then \emph{recovers} to $\sim$80\% by backbone layer 44, before declining again to $\sim$60\% at the decoder exit. This recovery suggests the backbone partially re-encodes style information from contextual representations, even having received collapsed audio tokens. CKA, by contrast, remains near 1.0 throughout — the most extreme CKA-probing dissociation in our study.

\subsection{Output-Level Collapse: Two categories}

Table~\ref{tab:tone} reports per-style tone perception accuracy. The gap between encoder probing ($\sim$82--85\%) and output accuracy (19.7\%--53.7\%) is the central empirical finding: the encoder is substantially richer than the output suggests.

\begin{table}[t]
\centering
\caption{Per-style tone perception accuracy under open-ended prompting. Chance = 14.29\%.}
\label{tab:tone}
\begin{tabular}{lccc}
\toprule
\textbf{Style} & \textbf{Qwen2-Audio} & \textbf{Qwen2.5-Omni} & \textbf{Chroma} \\
\midrule
confused    & 0.99\%  & 3.75\%  & 5.59\% \\
default     & 54.17\% & 66.64\% & 28.56\% \\
enunciated  & 46.64\% & 52.21\% & 36.84\% \\
happy       & 13.60\% & 43.95\% & 19.35\% \\
laughing    & 17.40\% & 85.11\% & 51.16\% \\
sad         & 3.75\%  & 30.35\% & 1.71\% \\
whisper     & 1.18\%  & 93.94\% & 14.76\% \\
\midrule
\textbf{Overall} & \textbf{19.68\%} & \textbf{53.70\%} & \textbf{22.57\%} \\
\bottomrule
\end{tabular}
\end{table}

\begin{figure}[t]
    \centering
    \includegraphics[width=0.95\textwidth]{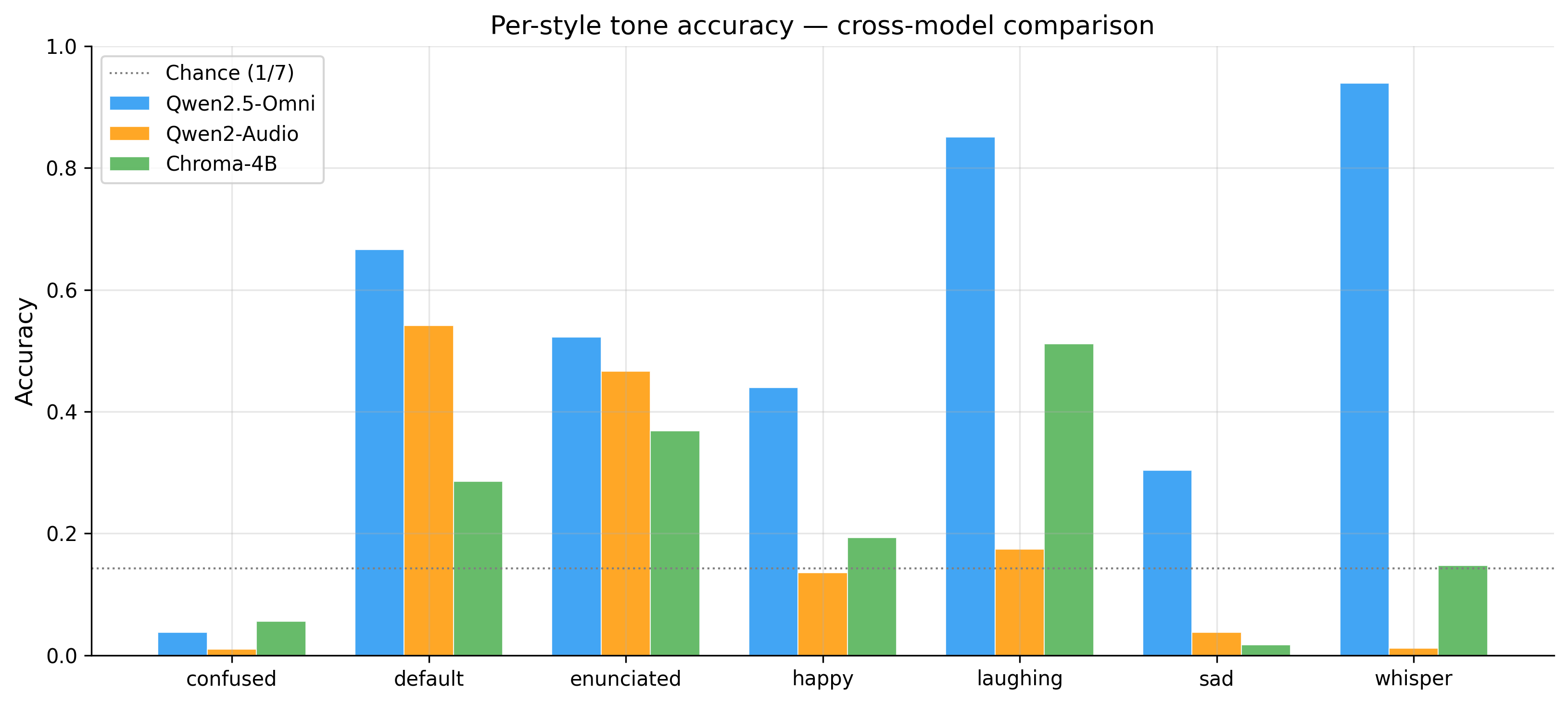}
    \caption{Per-style tone perception accuracy across all three models. Omni (blue) dominates on acoustically extreme styles (whisper, laughing). Qwen2-Audio (orange) performs best on clarity styles (default, enunciated). Chroma (green) partially recovers on laughing but fails on subtle styles. Chance = 14.3\% (dotted).}
    \label{fig:tone_cross}
\end{figure}

\begin{figure}[t]
    \centering
    \includegraphics[width=0.7\textwidth]{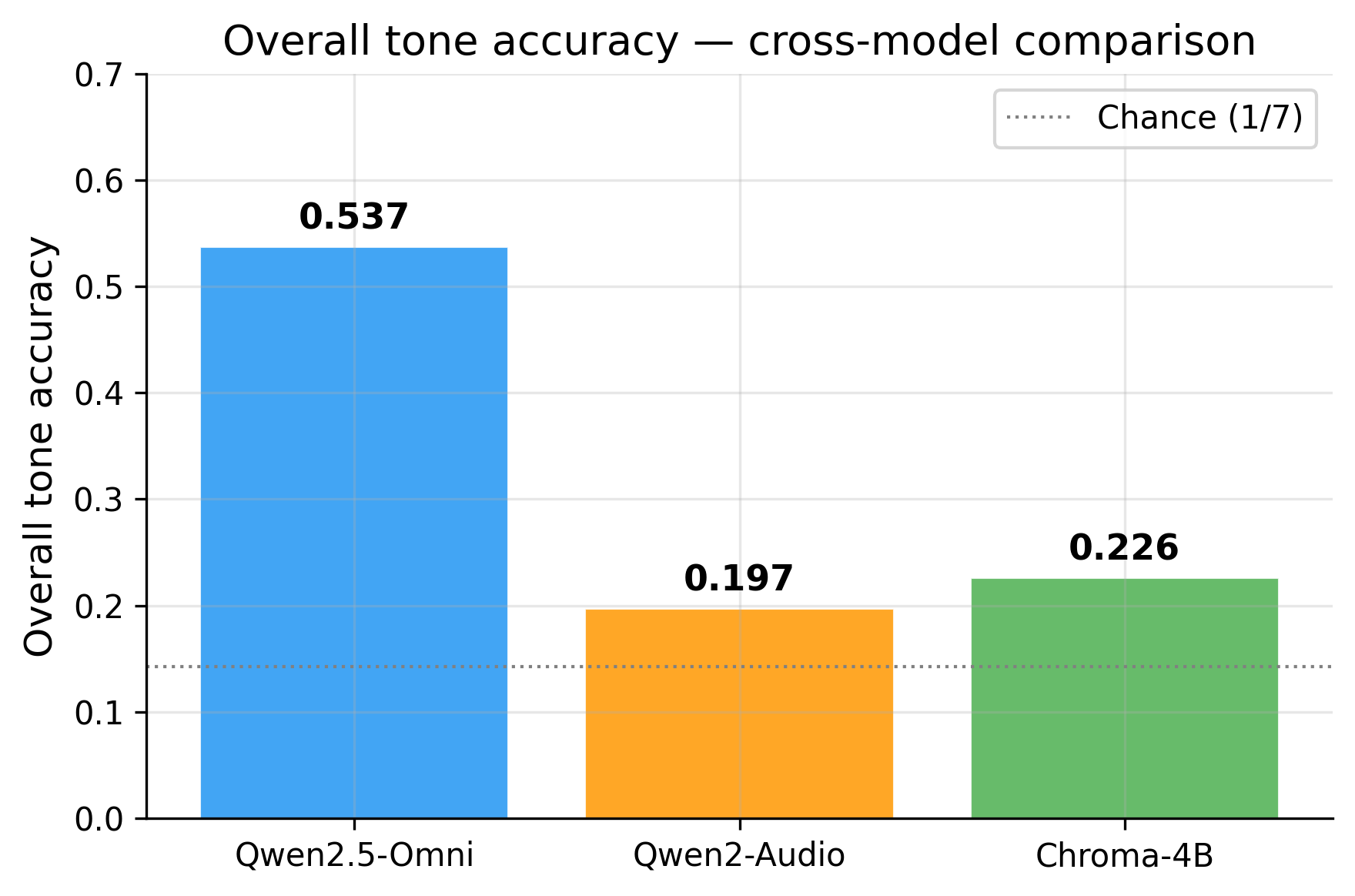}
    \caption{Overall tone perception accuracy. Omni (53.7\%) substantially outperforms Qwen2-Audio (19.7\%) and Chroma (22.6\%), all above chance (14.3\%).}
    \label{fig:tone_overall}
\end{figure}

Qwen2.5-Omni dominates on acoustically extreme styles (whisper: 93.94\%, laughing: 85.11\%). Qwen2-Audio performs best on speech clarity styles (default: 54.17\%, enunciated: 46.64\%). Chroma partially recovers on laughing (51.16\%) but fails on subtle styles — notably sad (1.71\%, below chance). All three models fail on confused ($<$7\%), confirming it as the hardest style for automatic evaluators. Figures~\ref{fig:tone_cross} and \ref{fig:tone_overall} illustrate these patterns. Omni's large advantage over the other two models is explained by its decoder behavior: Sections~\ref{sec:arch-analysis} and~\ref{sec:two-category} show that Omni's speech-generation training objective is the key factor that keeps acoustic style information alive through the decoder, whereas Qwen2-Audio and Chroma are trained only (or primarily) for text prediction and have no such incentive.

\subsection{Content-Prosody Leakage}

Table~\ref{tab:leakage} reports leakage statistics. Qwen2-Audio is 97.7\% content-driven — its tone predictions are almost entirely determined by what is said, not how it is said. Chroma is 84.4\% content-driven. Qwen2.5-Omni is only 27.2\% content-driven — it is genuinely listening to the acoustic signal. Because content ($x$) has substantially higher entropy than the 7-way style label ($e$), absolute leakage ratios are biased toward 1.0 for all three models; Appendix~\ref{app:leakage} directly tests this by recomputing leakage with content clustered into 7 semantic buckets to match style's cardinality, and confirms that the relative gap between Qwen2.5-Omni and Qwen2-Audio is robust to this bias even though the absolute values should not be read as calibrated probabilities.

\begin{table}[t]
\centering
\caption{Content-prosody leakage statistics. Leakage ratio $= \text{NMI}(\hat{t},x) / (\text{NMI}(\hat{t},x) + \text{NMI}(\hat{t},e))$. Higher = more content-driven. $\chi^2$ tests prediction independence from style (df=60).}
\label{tab:leakage}
\begin{tabular}{lrrr}
\toprule
\textbf{Metric} & \textbf{Qwen2-Audio} & \textbf{Qwen2.5-Omni} & \textbf{Chroma} \\
\midrule
Style accuracy                        & 19.68\%  & 53.70\% & 22.57\% \\
NMI($\hat{t}$, style)                 & 0.0056   & 0.3740  & 0.0469 \\
NMI($\hat{t}$, text)                  & 0.2335   & 0.1399  & 0.2542 \\
Leakage ratio                         & 0.977    & 0.272   & 0.844  \\
Mean agreement                        & 0.767    & 0.450   & 0.747  \\
\% perfect agreement                  & 30.53\%  & 0.59\%  & 19.93\% \\
$\chi^2$ (pred $\perp$ style)         & 208      & 19{,}519 & 2{,}343 \\
\bottomrule
\end{tabular}
\end{table}

The $\chi^2$ magnitude difference is striking: Qwen2.5-Omni's statistic is 94$\times$ larger than Qwen2-Audio's, reflecting the dramatic difference in acoustic sensitivity. The 0.59\% perfect agreement for Omni (vs.\ 30.53\% for Qwen2-Audio) means that almost no utterance receives the same tone prediction across all 7 speaking styles — Omni is consistently sensitive to acoustic variation.

\subsection{Qualitative Output Consistency}

Qualitative analysis confirms the two-category distinction mechanistically. For \textit{``Go to hell!''}: Chroma produces ``angry'' for all 7 speaking styles including whisper and laughing. Qwen2-Audio produces ``angry'' or ``disgusted'' for 6 of 7 styles. Qwen2.5-Omni differentiates: confused gets ``disgusted'', sad gets ``sad'', whisper gets ``whispering'', laughing gets ``disgusted'' — acoustically driven variation.

For semantically neutral content (\textit{``Monday there's gonna be haze but Tuesday look for thunderstorms''}): Chroma produces ``neutral'' for all 7 styles. Qwen2-Audio mirrors this. But Omni produces: confused $\to$ neutral, laughing $\to$ happy, sad $\to$ sleepy, whisper $\to$ whispering. Even on neutral content, Omni's predictions track acoustic style.

The ``sleepy'' label Omni assigns to sad speech is particularly informative: the model perceives low energy and reduced speech rate — genuine acoustic features of sad speech — but maps them to an imprecise vocabulary. This is a perception success with a vocabulary failure, distinct from the content-driven models' complete acoustic blindness.

Content-locked utterances (all models predict the same tone for all 7 styles) are exclusively those with explicit emotion words in the text: ``Happy reading!'', ``I am really, really happy.'' This validates the metric — lexically explicit content predictably collapses all models to content-driven prediction.

%% file: sections/discussion.tex
\section{Discussion}

\subsection{The Encoder-Output Gap}

Across all models, we observe a consistent gap between encoder-level probing accuracy ($\sim$82--85\%) and output-level tone perception (19.7\%--53.7\%). This gap is reproducible across methods, models, and speaking styles, indicating that it is a structural property of the ALM pipeline rather than a model-specific artifact. Leakage analysis shows that the underlying causes differ by model. For Qwen2-Audio and Chroma, the decoder ignores acoustic representations in favor of semantic priors. For Omni, the gap is smaller and arises from a different issue: the decoder uses acoustic information, but the output vocabulary is misaligned with Expresso's style taxonomy.

\subsection{Architectural Analysis}
\label{sec:arch-analysis}

Three factors govern paralinguistic preservation:

\textbf{Projector type.} Qwen2-Audio's linear projector reshapes representation geometry without removing information, resulting in flat decoder probing. In contrast, Omni's MLP projector introduces gradual information erosion.

\textbf{Training objective.} Omni's speech generation objective encourages preservation of acoustic style through the decoder, explaining its acoustic-driven behavior despite sharing the same encoder as Qwen2-Audio. Chroma's voice cloning objective preserves style for synthesis but not for text prediction, leading to a sharp collapse at the audio-to-language boundary.

\textbf{Component boundaries.} In Chroma, style collapses completely at the thinker audio to thinker LM boundary (layer 32), so the language model never receives style-differentiated inputs. This explains both its high leakage ratio and flat CKA. The subsequent recovery in probing suggests that the Llama3-based backbone partially reconstructs style from context.

\subsection{The Two-category Hypothesis}
\label{sec:two-category}

Our results suggest two modes of output behavior. In content-driven models, the decoder relies on learned associations between text and emotion, overriding acoustic signals. In acoustic-driven models, the decoder conditions directly on acoustic representations. The primary factor determining this behavior is the training objective. Models trained only for text prediction (Qwen2-Audio, Chroma thinker LM) have no incentive to preserve acoustic information, while models trained for speech generation (Omni) maintain it through the decoder.

\subsection{Closing the Encoder-Output Gap}
\label{sec:grl}

To test whether the gap can be reduced at the projector boundary, we applied Gradient Reversal Layer (GRL) training \citep{ganin2016domain} to Qwen2-Audio's projector, the only interface between the audio encoder and the LLM. The projector is a single affine transformation ($W \in \mathbb{R}^{4096 \times 1280}$), making it a minimal intervention point. We trained only the projector ($\sim$5M parameters) using two objectives: a style classification head to maximize style decodability and an adversarial content head with gradient reversal to suppress content information. GRL reduced content signal (NMI($\hat{t}, x$): $0.234 \to 0.203$) but also reduced style information (NMI($\hat{t}, e$): $0.006 \to 0.004$), leaving the leakage ratio unchanged ($0.977 \to 0.979$). This result is expected: a linear transformation cannot disentangle linearly entangled signals. If style and content share dimensions in the encoder space, no linear map can separate them. Effective disentanglement therefore requires either a nonlinear projector or training the encoder to produce disentangled representations.

\subsection{Implications for Model Design}
\label{sec:design-implications}

Two findings are directly actionable. First, the GRL result (Section~\ref{sec:grl}) shows style and content are linearly entangled in encoder space, ruling out linear projector-level fixes and pointing toward nonlinear projectors or encoder-level contrastive objectives. Second, Qwen2-Audio vs.\ Qwen2.5-Omni shows architecture is not the bottleneck: nearly identical encoders yield sharply different output behavior (leakage ratio 97.7\% vs.\ 27.2\%) purely from Omni's speech-generation objective. Closing the encoder-output gap thus looks like a training-objective and representation-geometry problem, not a scaling or restructuring one. This motivates three directions:

\textbf{Training objective.} Auxiliary objectives, such as predicting speaking style from audio tokens during instruction tuning, could reduce the gap without architectural change, mirroring Omni's speech-generation objective.

\textbf{Targeted interventions.} Layer-wise probing localizes where paralinguistic information is accessible, enabling targeted follow-up methods beyond the general-purpose ones used here, such as activation patching, value zeroing, and representation steering \citep{zou2023representation}; we view these as natural next steps our localization makes tractable, not claims we substantiate here.

\textbf{Contrastive objectives.} Motivated by the GRL result, we propose training on paired same-content, different-style utterances $(a(x, e_i, s), a(x, e_j, s)) \in \mathcal{P}$ with a nonlinear or contrastive objective, minimizing NMI($\hat{t}, x$) and maximizing NMI($\hat{t}, e$) to directly encourage acoustic sensitivity.

\subsection{Limitations}
\label{sec:limitations}

Our analysis uses four Expresso speakers; LOSO prevents speaker leakage but generalization to larger populations is not guaranteed. Mean pooling discards temporal dynamics that may matter for styles like laughing, though the consistent $\sim$82--85\% probing peaks (Section~\ref{sec:results}) suggest most linearly accessible signal survives it. The leakage metric's absolute values are biased by the content/style entropy asymmetry (Section~\ref{sec:results}); relative comparisons remain valid under this bias. Linear probes could in principle underestimate information at layers with shifting geometry; we tested this with an MLP probe at each model's key boundary layers (Appendix~\ref{app:mlp}). Across 19 boundary layers, MLP accuracy stayed within $-4.1$ to $+0.4$ points of linear (mean $-1.0$), with no layer meaningfully higher, indicating our linear probes are near the decodability ceiling rather than underestimating it. We restrict our analysis to speaking style; other paralinguistic attributes (speaker identity, accent, health-related signals) are out of scope. We study only text outputs, since Whisper and Qwen2-Audio do not generate speech; whether Omni or Chroma additionally leverage prosody when generating speech is an open question. Finally, the leakage metric assumes consistent tone mapping across models' differing output vocabularies, and our toolkit identifies \emph{where} information is lost but not, by itself, \emph{why}; Section~\ref{sec:design-implications} outlines causal follow-ups this localization motivates.

%% file: sections/conclusion.tex
\section{Conclusion}

We have traced paralinguistic information through the full pipeline of four audio-language models using four complementary methods. All models encode speaking style strongly in the late audio encoder ($\sim$82--85\% probe accuracy), but this information is systematically underutilized at the output level. The projector acts as a geometry transformer rather than an information bottleneck. Decoders either preserve (Qwen2-Audio), erode (Omni), or collapse (Chroma) style information depending on architecture and training objective. At the output level, models cluster into content-driven (leakage ratio $>$84\%) and acoustic-driven (27.2\%) categories, with training objective as the primary determinant. These findings motivate future work on paralinguistic-aware training objectives and architectural designs that preserve expressive information across the audio-to-language boundary.

%% file: sections/appendix.tex
\appendix

\section{CKA Implementation Details}
\label{app:cka}

All CKA computations use linear kernels with mean-centering. For paired clips $(a_1, a_2)$, we extract hidden states $H_1, H_2 \in \mathbb{R}^{T \times d}$ and mean-pool over $T$ to obtain $h_1, h_2 \in \mathbb{R}^d$. We aggregate over all pairs of a given style combination to form matrices $X, Y \in \mathbb{R}^{n \times d}$. A NaN guard excludes pairs where the denominator falls below $10^{-10}$ (degenerate representations). All 31,896 pairs are used for aggregation.

\section{Probe Training Details}
\label{app:probe}

We use scikit-learn's \texttt{LogisticRegression} with L2 regularization ($C=0.1$), \texttt{max\_iter=500}, \texttt{solver=saga}, \texttt{tol=1e-3}, and \texttt{class\_weight=balanced}. The SAGA solver was selected for faster convergence on high-dimensional hidden states. Four LOSO folds (one per speaker). Features standardized per fold using training set statistics. Results reported as mean $\pm$ std across folds.

\section{Nonlinear (MLP) Probing at Boundary Layers}
\label{app:mlp}

To test whether our linear probes underestimate the paralinguistic information available at layers where representational geometry shifts substantially (e.g.\ across the projector or backbone boundaries), we trained a nonlinear probe, a single-hidden-layer MLP (scikit-learn \texttt{MLPClassifier}, hidden width 256, L2 $\alpha=10^{-3}$, \texttt{max\_iter=300}, early stopping), under the same LOSO protocol and feature standardization as the linear probes, restricted to each model's key boundary layers: the late-encoder accuracy peak, the projector (where applicable), and the decoder entry, middle, and exit layers (for Chroma: the thinker-audio peak, backbone entry, backbone recovery point, and decoder entry/exit). Table~\ref{tab:mlp_probe} compares linear and MLP accuracy at each of these 19 layers across the four models.

\begin{table}[h]
\centering
\caption{Linear vs.\ nonlinear (MLP) probe accuracy (LOSO) at each model's key boundary layers. $\Delta$ = MLP $-$ linear. No layer shows a meaningfully positive $\Delta$; the linear probe accuracy is close to the ceiling of what is decodable at these layers.}
\label{tab:mlp_probe}
\begin{tabular}{llrrr}
\toprule
\textbf{Model} & \textbf{Layer} & \textbf{Linear} & \textbf{MLP} & \textbf{$\Delta$} \\
\midrule
Qwen2-Audio & enc\_20 (encoder peak)     & 0.8196 & 0.8020 & $-$0.018 \\
            & projector                  & 0.7851 & 0.7814 & $-$0.004 \\
            & dec\_0 (decoder entry)     & 0.7851 & 0.7814 & $-$0.004 \\
            & dec\_16 (decoder mid)      & 0.7905 & 0.7834 & $-$0.007 \\
            & dec\_32 (decoder exit)     & 0.7872 & 0.7792 & $-$0.008 \\
\midrule
Qwen2.5-Omni & enc\_22 (encoder peak)    & 0.8534 & 0.8483 & $-$0.005 \\
             & projector                 & 0.8134 & 0.8177 & $+$0.004 \\
             & dec\_0 (decoder entry)    & 0.8123 & 0.8161 & $+$0.004 \\
             & dec\_13 (decoder mid)     & 0.7842 & 0.7707 & $-$0.014 \\
             & dec\_27 (decoder exit)    & 0.7359 & 0.7363 & $+$0.000 \\
\midrule
Chroma-4B & thinker\_audio\_20 (peak)    & 0.8531 & 0.8507 & $-$0.002 \\
          & backbone\_0 (entry)          & 0.5723 & 0.5566 & $-$0.016 \\
          & backbone\_12 (recovery)      & 0.7985 & 0.7894 & $-$0.009 \\
          & decoder\_0                   & 0.6598 & 0.6186 & $-$0.041 \\
          & decoder\_3 (exit)            & 0.5953 & 0.5696 & $-$0.026 \\
\midrule
Whisper-large-v2 & enc\_0                & 0.5001 & 0.4787 & $-$0.021 \\
                 & enc\_15               & 0.7811 & 0.7820 & $+$0.001 \\
                 & enc\_24 (peak)        & 0.8346 & 0.8220 & $-$0.013 \\
                 & enc\_31 (final)       & 0.8160 & 0.8035 & $-$0.013 \\
\bottomrule
\end{tabular}
\end{table}

Across all 19 boundary layers, the MLP-linear difference ranges from $-4.1$ to $+0.4$ points, with a mean of $-1.0$ points, and no layer shows a substantively higher nonlinear accuracy. This supports treating our linear probe accuracies as close to the ceiling of linearly \emph{and} nonlinearly decodable information at these layers, rather than as an underestimate caused by information being encoded in a nonlinearly accessible form.

\section{Tone Perception Prompt and Keyword Mapping}
\label{app:prompt}

\textbf{Prompt:} \textit{``Reply in English only. What is the person saying? Describe the tone briefly (e.g.\ neutral, happy, sad, angry, whispering, confused, laughing).''}

\textbf{Tone bucket mapping} (case-insensitive, first match wins):
\begin{itemize}
    \item \textit{angry}: angry, frustrated, aggressive, forceful
    \item \textit{happy}: happy, cheerful, joyful, amused, excited, light-hearted
    \item \textit{sad}: sad, somber, sorrowful, melancholic
    \item \textit{neutral}: neutral, informative, matter-of-fact, calm, straightforward
    \item \textit{whisper}: whisper, whispering
    \item \textit{laughing}: laugh, laughing, giggling
    \item \textit{confused}: confused, puzzled, bewildered
    \item \textit{disgusted}: disgusted, disgust
    \item \textit{surprised}: surprised, astonished, shocked
    \item \textit{sleepy}: sleepy, drowsy, tired
    \item \textit{other}: no match
\end{itemize}

\section{Layer Coverage by Model}
\label{app:layers}

\begin{table}[h]
\centering
\caption{Layer coverage per model.}
\begin{tabular}{lrrrr}
\toprule
\textbf{Model} & \textbf{Encoder} & \textbf{Projector} & \textbf{Decoder} & \textbf{Total} \\
\midrule
Whisper-large-v2    & 32 & —  & —    & 32  \\
Qwen2-Audio-7B      & 33 & 1  & 33   & 67  \\
Qwen2.5-Omni-7B     & 32 & 1  & 28   & 61  \\
Chroma-4B           & 32 & —  & 16+4 & 52$^\dagger$ \\
\bottomrule
\end{tabular}

\vspace{0.5em}
\parbox{0.95\textwidth}{\footnotesize $^\dagger$Chroma's full architecture is 32 (thinker audio) + 36 (thinker LM) + 16 (backbone) + 4 (decoder) = 88 layers. The thinker LM (36 layers) is excluded: CKA shows complete style collapse at the thinker audio-to-LM boundary ($\sim$0.98--1.0 throughout). Probed layers: thinker audio (32) + backbone (16) + decoder (4) = 52, renumbered sequentially.}
\end{table}

\clearpage
\section{Per-Pair CKA Trajectories}
\label{app:cka_fig}

\begin{figure}[!ht]
    \centering
    \begin{subfigure}[b]{0.48\textwidth}
        \includegraphics[width=\textwidth]{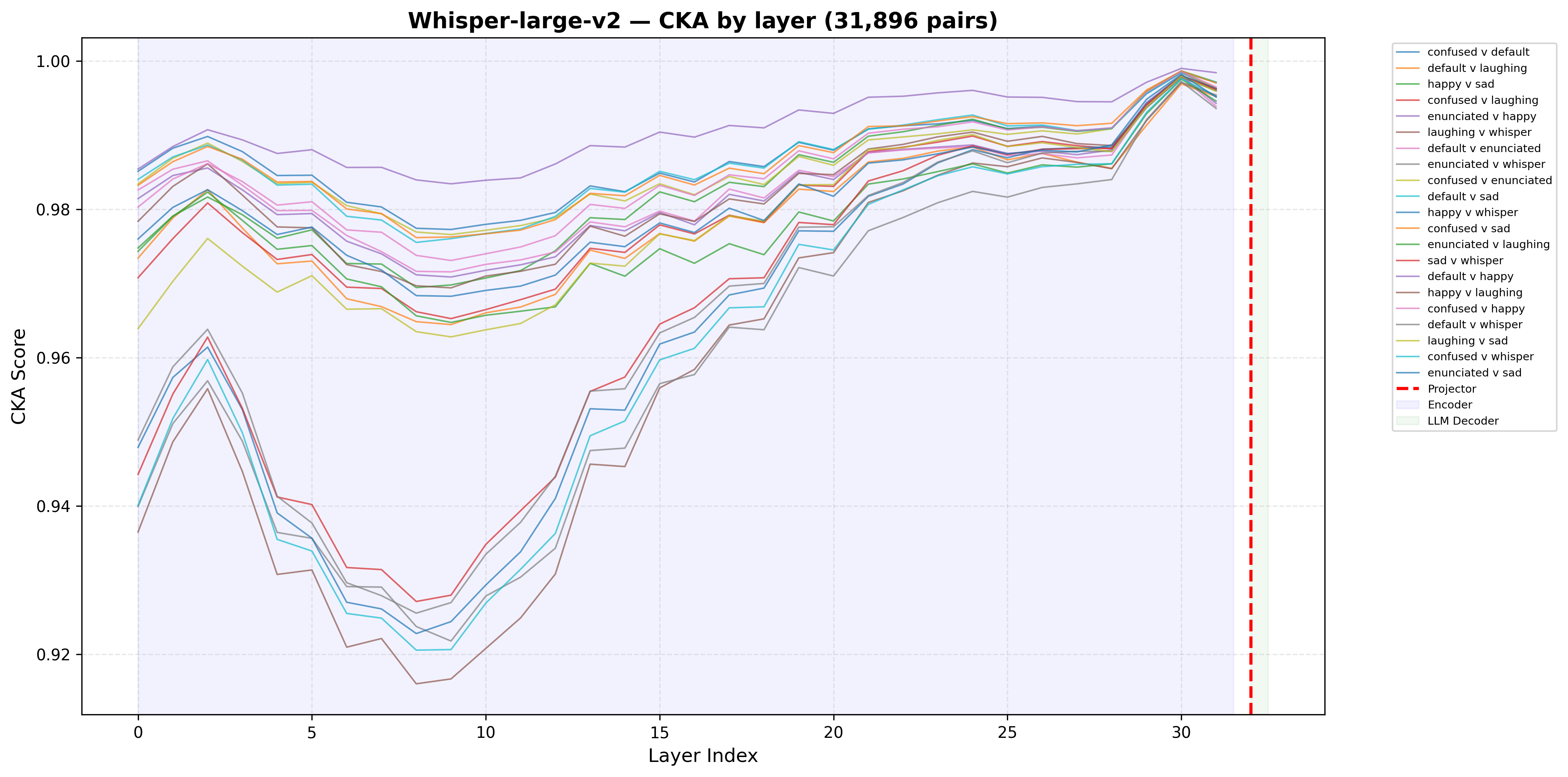}
        \caption{Whisper-large-v2}
    \end{subfigure}
    \hfill
    \begin{subfigure}[b]{0.48\textwidth}
        \includegraphics[width=\textwidth]{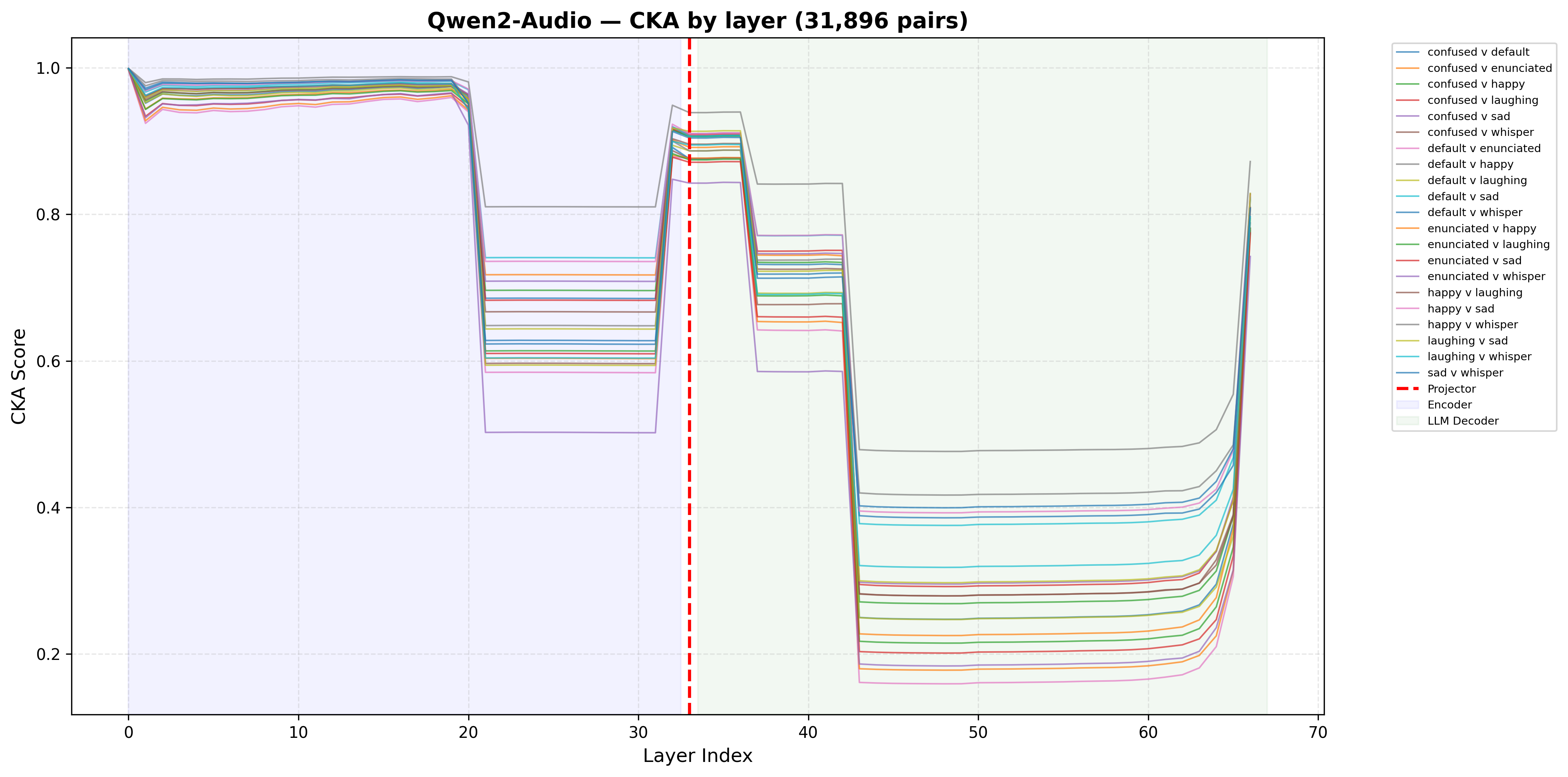}
        \caption{Qwen2-Audio-7B-Instruct}
    \end{subfigure}
    \vspace{0.5em}
    \begin{subfigure}[b]{0.48\textwidth}
        \includegraphics[width=\textwidth]{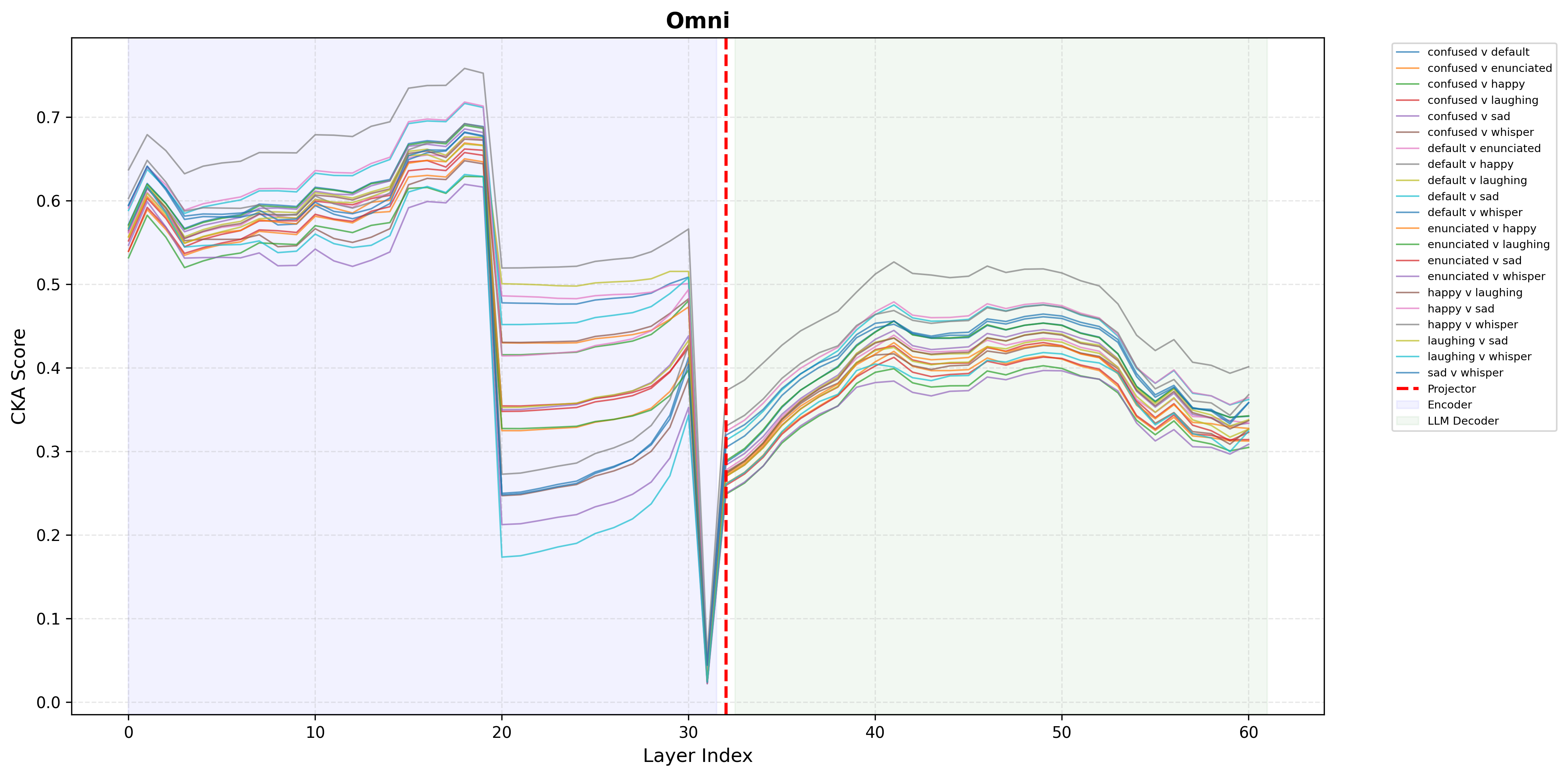}
        \caption{Qwen2.5-Omni-7B}
    \end{subfigure}
    \hfill
    \begin{subfigure}[b]{0.48\textwidth}
        \includegraphics[width=\textwidth]{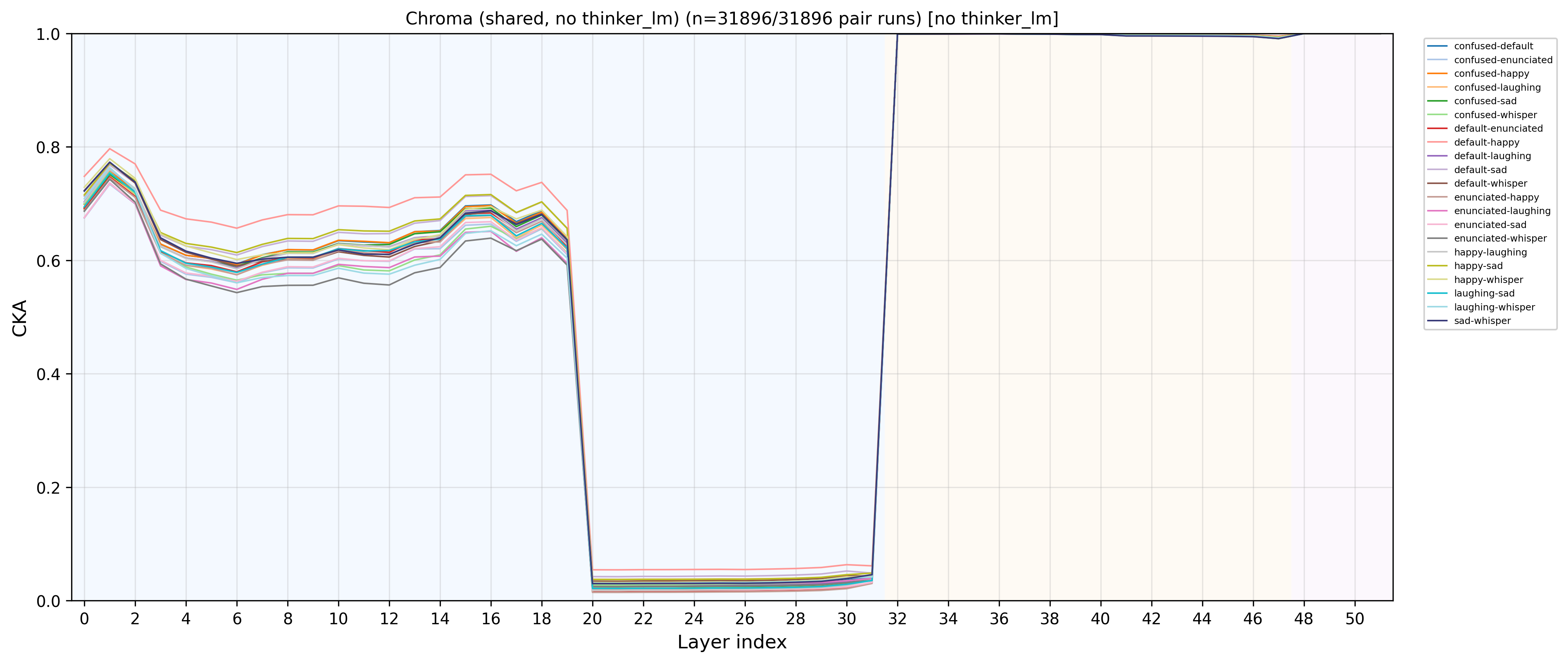}
        \caption{Chroma-4B}
    \end{subfigure}
    \caption{CKA by layer for all $\binom{7}{2}=21$ style pairs (one line per pair), shown here at full per-pair granularity; Section~\ref{sec:results} summarizes the aggregate trend. Red dashed line marks the projector boundary. Whisper shows near-uniform geometric similarity (high CKA) across all pairs. All three ALMs show late-encoder style separation (decreasing CKA), consistently across pairs. Chroma collapses all style differences at the backbone boundary (layer 32).}
    \label{fig:cka_all}
\end{figure}

\section{Content-Prosody Leakage: Additional Details}
\label{app:leakage}

Agreement score is computed per unique utterance text across all 7 speaking styles. The leakage ratio uses normalized mutual information (NMI) to account for differences in label entropy. Chi-squared tests use the full contingency table of predicted tone buckets vs.\ style labels.

As noted in Section~\ref{sec:results}, content ($x$, one value per unique utterance) has substantially higher entropy than the 7-class style label ($e$), which biases absolute leakage ratios toward 1.0 for all models. To directly test whether this bias distorts the comparison between models, we recomputed the leakage ratio with content mapped to 7 semantic clusters (TF-IDF features, $k$-means, $k=7$ to match $|\mathcal{E}|=7$) instead of raw per-utterance text identity. Table~\ref{tab:leakage_balanced} reports both versions. Balancing granularity lowers all three leakage ratios in absolute terms, as expected, but preserves the relative ordering and the size of the gap between models: Qwen2-Audio remains the most content-driven (0.787), Chroma intermediate (0.318), and Omni the most acoustic-driven (0.020), the same ordering as with raw text. This confirms that the entropy asymmetry inflates absolute leakage values but does not drive the cross-model comparison that the paper's claims rest on.

\begin{table}[h]
\centering
\caption{Leakage ratio under raw per-utterance text content vs.\ content clustered into 7 semantic buckets (matching the 7-way style label). Ordering across models is unchanged.}
\label{tab:leakage_balanced}
\begin{tabular}{lrrr}
\toprule
\textbf{Metric} & \textbf{Qwen2-Audio} & \textbf{Qwen2.5-Omni} & \textbf{Chroma} \\
\midrule
Leakage ratio (raw text)        & 0.977 & 0.272 & 0.844 \\
Leakage ratio (7-cluster text)  & 0.787 & 0.020 & 0.318 \\
\bottomrule
\end{tabular}
\end{table}

Content-locked utterances (perfect agreement across all 3 models, all 7 styles, $n=7$) are exclusively those with explicit emotion words in the text: \textit{``Happy reading!''}, \textit{``I am really, really happy.''}, and similar. This validates the metric — lexically explicit emotion content predictably collapses all models to content-driven prediction.
\section{GRL Intervention: Training Details}
\label{app:grl}

To test whether the encoder-output gap identified in Section~\ref{sec:results}
could be addressed by intervening at the projector boundary, we trained
Qwen2-Audio's multimodal projector with a Gradient Reversal Layer (GRL)
objective \citep{ganin2016domain}.

\textbf{Architecture.} All model weights were frozen except the multimodal
projector ($W \in \mathbb{R}^{4096 \times 1280}$, bias=True, $\sim$5M
parameters). Two lightweight linear heads were attached to the mean-pooled
projector output $\bar{h}_{proj} \in \mathbb{R}^{4096}$: a style head
($\mathbb{R}^{4096} \to \mathbb{R}^7$) and a content head
($\mathbb{R}^{4096} \to \mathbb{R}^{20}$).

\textbf{Content labels.} Utterance texts were clustered into 20 semantic
clusters via TF-IDF + KMeans, providing a tractable 20-way content
classification target for the adversarial head.

\textbf{Loss function.} For each batch:
\begin{equation}
\mathcal{L} = \mathcal{L}_\text{style}(h_{proj},\, e)
            + \lambda \cdot \mathcal{L}_\text{content}(\text{GRL}(h_{proj}),\, c)
\end{equation}
where $e$ is the style label, $c$ is the text cluster label, and $\lambda$
follows the schedule of \citet{ganin2016domain}:
\begin{equation}
\lambda(t) = \frac{2}{1 + e^{-\gamma t / T}} - 1, \quad \gamma = 10
\end{equation}
ramping from 0 to 1 over training. No language modeling loss was used,
avoiding the generation collapse observed in a preliminary run.

\textbf{Training.} 5 epochs, AdamW ($\eta$=1e-4, weight decay=0.01),
cosine schedule with 5\% warmup, batch size 8, 8 * V100-32GB.
Dataset: 8,511 training clips / 2,123 validation clips from Expresso,
stratified by style, 80/20 split.

\textbf{Results.} Style head validation accuracy reached 93.3\% by epoch 4,
confirming that projector representations contain sufficient information for
near-perfect style classification under direct supervision. Content head
accuracy stabilized at 14--22\% (vs.\ 5\% chance), indicating partial but
incomplete content suppression. The best checkpoint (epoch 2, lowest content
accuracy 14.2\%) was used for generation evaluation.
Generation outputs under the blind evaluation prompt showed proportional
suppression of both content and style signal, with the leakage ratio
unchanged (Table~\ref{tab:grl}).

\begin{table}[h]
\centering
\caption{GRL intervention results. Epoch 2 checkpoint (lowest content accuracy).}
\label{tab:grl}
\begin{tabular}{lrrl}
\toprule
\textbf{Metric} & \textbf{Baseline} & \textbf{GRL} & \textbf{$\Delta$} \\
\midrule
Style accuracy       & 19.68\% & 18.01\% & $-$1.67\% \\
Leakage ratio        & 0.977   & 0.979   & $+$0.002  \\
NMI($\hat{t}$, style) & 0.006  & 0.004   & $-$0.002  \\
NMI($\hat{t}$, text)  & 0.234  & 0.203   & $-$0.031  \\
Whisper accuracy     & 1.18\%  & 0.13\%  & $-$1.05\% \\
Laughing accuracy    & 17.40\% & 0.44\%  & $-$16.96\% \\
\bottomrule
\end{tabular}
\end{table}